# Efficient computation of redundancy matrices for moderately redundant truss and frame structures

🅳Anton TKACHUK[1], Tim KRAKE[2,3], 🅳Jan GADE[4], and 🅳Malte von SCHEVEN[4]

[1] Karlstad University, Department of Engineering and Physics, SE-658 88 Karlstad, Sweden
[2] University of Stuttgart, Visualization Research Center (VISUS)
[3] Hochschule der Medien, Stuttgart
[4] University of Stuttgart, Institute for Structural Mechanics

Large statically indeterminate truss and frame structures exhibit complex load-bearing behavior, and redundancy matrices are helpful for their analysis and design. Depending on the task, the full redundancy matrix or only its diagonal entries are required. The standard computation procedure has a high computational effort. Many structures fall in the category of moderately redundant, i.e., the ratio of the statical indeterminacy to the number of all load-carrying modes of all elements is less one half. This paper proposes a closed-form expression for redundancy contributions that is computationally efficient for moderately redundant systems. The expression is derived via a factorization of the redundancy matrix that is based on singular value decomposition. Several examples illustrate the behavior of the method for increasing size of systems and, where applicable, for increasing degree of statical indeterminacy.

**Keywords:** redundancy matrix, fast computation, matrix decomposition, truss structures, frame structures

## 1 Introduction

Information about the system-inherent properties of structures provides valuable insights into their behavior. For a usable spatially discrete truss or frame structure, the degree of statical indeterminacy is a widely used system-inherent property in elastostatics. It is equal to the difference between the number of unknown force and moment quantities and the number of linearly independent equilibrium equations. The statical indeterminacy is commonly considered as one aggregated integer number characterizing the number of redundant constraints within the entire structure. Unfortunately, statical indeterminacy does not describe the contribution of each member of the structure to load-carrying properties. Information about the spatial distribution of the degree of statical indeterminacy, as well as information about different load-carrying mechanisms, can be obtained from the redundancy matrix. This information shows the variety of load paths in the structure, which can be later used in the analysis and design of the structure. Particularly, the importance of individual members can be concluded from diagonal entries of the redundancy matrix, also called here distribution of statical indeterminacy.

A concept of redundancy in statically indeterminate truss and frame structures was proposed on the basis of the analogy between Gaussian adjustment calculus in geodesy and structural mechanics (Linkwitz 1961; Bahndorf 1991; Ströbel 1997). Recently, the idea of the redundancy matrix was extended to kinematically indeterminate structures (Tibert 2005; Zhou et al. 2015; Chen et al. 2018) and to continuum mechanical framework (Gade et al. 2021). A more extensive literature overview on the calculation and application of the redundancy matrix can be found in the work by von Scheven et al. (2021). The major applications of redundancy matrix in the design of structures aim primarily at the reliability and robustness of structures (Frangopol and Curley 1987; Pandey and Barai 1997; Kou et al. 2017; Spyridis and Strauss 2020) and the quantification of imperfection sensitivity (Eriksson and Tibert 2006; Ströbel and Singer 2008). More recent applications include adaptability, actuator placement, and optimized control in adaptive structures (Wagner et al. 2018; Geiger et al. 2020; Maierhofer and Menges 2019). A detailed discussion how the redundancy matrix can be used in the design and analysis of structures is beyond the scope of this paper and can be found in the references given above.

During the design process of structures, many different variants are usually studied. In this case, the redundancy distribution in the structure can be used as an indicator of the internal constraint or robustness of the different variants. These values can be either used as direct feedback to interactive design changes by an engineer or in an automatic optimization process (Bahndorf 1991). In both cases, a fast calculation of the redundancy matrix is essential. Recently, an algorithm for efficient updates of redundancy matrices in truss and frame structures is proposed in Krake et al. (2022). Therein, in distinction to the present contribution, generic algebraic formulations due to various modifications like adding, removing, and exchanging elements are employed. However, especially for larger structures, the calculation of the redundancy matrix itself is time-consuming as it involves the inverse of a matrix whose dimensionality is given by the number of degrees of freedom in the system. Therefore, this paper addresses the problem of efficient computation of the redundancy matrix or only its diagonal entries for large truss and frame structures, enabling rapid feedback to an engineer or faster optimization loops. In detail, we propose

- a decomposition of the redundancy matrix based on a singular value decomposition (SVD),
- application of this decomposition for efficient computation of the redundancy matrix or its diagonal entries,
- a rough range where the proposed formula provides a significant speedup,
- a set of benchmarks to investigate speedup.

In Section 2, all relevant aspects of matrix structural analysis and the definition of the redundancy matrix are provided. In Section 3, the SVD-based decomposition of the redundancy matrix is introduced and applied to derive efficient algorithms for the computation of the entire redundancy matrix and its diagonal entries only. The examples in Section 4 demonstrate the application of the proposed efficient computation formulations for different types of structures. Here, the dependency of the speedup on the relative degree of statical indeterminacy and the problem size is discussed. The paper concludes with a summary and an outlook.

## 2  Background

In this section, a brief summary of matrix structural analysis is given, including the equilibrium, material, and redundancy matrices and their relevant properties.

### 2.1  Matrix structural analysis

Since the 1960s, static analysis for discrete models of spatial truss and frame structures purified the matrix notation (Argyris and Scharpf 1969; Przemieniecki 1968), which is shortly described for displacement-based version in the following.

Given is a discrete structural model consisting of $n$ degrees of freedom, $n_n$ nodes, and $n_e$ elements. Elements are distinguished by nodal degrees of freedom and the number of load-carrying modes $n_m$. In the case of a pin-joined plane or spatial truss structure, each element possesses one load-carrying mode (axial tension/compression), i.e., $n_m = 1$. In the case of a frame structure, the number of load-carrying modes is $n_m = 6$ for a spatial and $n_m = 3$ for plane 2-node beam element. In general, a mixture of truss and beam elements is allowed in a structural model, i.e., $n_m$ may vary between the elements. Therefore, the index $n_q$ is introduced that defines the number of all load-carrying modes of all elements in the model ($n_q = n_m n_e$ in case of a pure truss or beam model).

Compatibility equations relate generalized displacements $\mathbf{d} \in \mathbb{R}^n$ to the generalized elastic strains $\mathbf{e}_{el} \in \mathbb{R}^{n_q}$ and the generalized pre-strains $\mathbf{e}_0 \in \mathbb{R}^{n_q}$

$$\mathbf{e}_{el} = \mathbf{A}\mathbf{d} - \mathbf{e}_0. \tag{1}$$

Here, $\mathbf{A} \in \mathbb{R}^{n_q \times n}$ denotes the compatibility matrix. Material equations relate the generalized elastic strains to the generalized stress resultants $\mathbf{s} \in \mathbb{R}^{n_q}$ according to

$$\mathbf{s} = \mathbf{C}\mathbf{e}_{el}, \tag{2}$$

with $\mathbf{C} \in \mathbb{R}^{n_q \times n_q}$ being the material matrix. It is diagonal with positive entries for truss and beam elements, as described in von Scheven et al. (2021) and shown for a 2-node straight 3D beam in Appendix A. The equilibrium equations relate the generalized stress resultants to the external loads $\mathbf{f} \in \mathbb{R}^n$

$$\mathbf{A}^\top \mathbf{s} = \mathbf{f}, \tag{3}$$

using the equilibrium matrix $\mathbf{A}^\top \in \mathbb{R}^{n \times n_q}$. The irreducible form of the structural equations w.r.t the vector of generalized displacements

$$\mathbf{A}^\top \mathbf{C}\mathbf{A}\mathbf{d} = \mathbf{f} + \mathbf{A}^\top \mathbf{C}\mathbf{e}_0 \tag{4}$$

satisfies these three sets of Equations (1) to (3). The square matrix

$$\mathbf{K} = \mathbf{A}^\top \mathbf{C}\mathbf{A} \in \mathbb{R}^{n \times n} \tag{5}$$

is called the elastic stiffness matrix. It is symmetric by construction due to the symmetry of $\mathbf{C}$ following from its diagonal shape.

Several formal conditions on dimensions and indices of the model are assumed throughout the paper. First, the considered structures are statically indeterminate with a degree of statical indeterminacy $n_s = n_q - \text{rank}(\mathbf{A}^\top)$. This condition ensures a nontrivial distribution of redundancy in the structures. Second, the structures are kinematically determinate or in other words, usable, i.e., $\text{rank}(\mathbf{A}) = n$ (Pellegrino and Calladine 1986; Pellegrino 1990; Pellegrino 1993). This condition ensures full rank of the elastic stiffness matrix $\mathbf{K}$.

## 2.2 Redundancy distribution

Using the notations and assumptions of the previous subsection, the concept of the redundancy (Bahndorf 1991; Ströbel 1997; von Scheven et al. 2021) is described in the following.

The redundancy is a property independent of the external loads; thus, $\mathbf{f} = \mathbf{0}$ is assumed. Elimination of the generalized displacement vector $\mathbf{d}$ from the compatibility Equation (1) using Equation (4) yields

$$-\mathbf{e}_{el} = \left(\mathbf{I} - \mathbf{A}\mathbf{K}^{-1}\mathbf{A}^\top \mathbf{C}\right)\mathbf{e}_0 = \mathbf{R}\mathbf{e}_0, \tag{6}$$

with the redundancy matrix

$$\mathbf{R} = \mathbf{I} - \mathbf{A}\mathbf{K}^{-1}\mathbf{A}^\top \mathbf{C} \in \mathbb{R}^{n_q \times n_q}. \tag{7}$$

It is a square matrix with a dimension equal to the number of all load-carrying modes $n_q$. The main-diagonal entries $r_{ii}$ provide the distribution of the degree of statical indeterminacy of the structure (Bahndorf 1991; Ströbel 1997) between the load-carrying modes such that $\sum_i^{n_q} r_{ii} = n_s$. The defining property of the redundancy matrix $\mathbf{R}$ in Equation (6) is that it maps the generalized pre-strains $\mathbf{e}_0$ onto negative elastic strains, i.e., it reveals the incompatible, stress-inducing part of the action of $\mathbf{e}_0$. In the case of a statically determinate structure, $\mathbf{R}$ is a zero matrix and any $\mathbf{e}_0$ leads to $\mathbf{e}_{el} = \mathbf{0}$.

The computational complexity for the calculation of the whole redundancy matrix is dominated by the inverse of the stiffness matrix and two matrix-matrix multiplications. The former operation has cost depending on the bandwidth of the stiffness matrix, but it is limited from above by the value for a dense matrix of the same size $O(n^3)$. The latter operation has cost $O(n \cdot n_q^2)$. Since $n$ is typically proportional to $n_q$, the complexity scales as the third power of the problem size.

If only the distribution of the degree of statical indeterminacy in the structure is required, the calculation can be performed only for the main-diagonal entries of the redundancy matrix in Equation (7). Defining $\mathbf{a}_i \in \mathbb{R}^{1 \times n}$ as the $i$-th row of the compatibility matrix $\mathbf{A}$, the following canonical formula can be used:

$$r_{ii} = 1 - \mathbf{a}_i \mathbf{K}^{-1} (c_{ii} \mathbf{a}_i)^\top \tag{8}$$

This simple and straightforward modification of the definition of the redundancy matrix in Equation (7) makes use of simple linear algebra manipulations only to compute the diagonal entries. Therefore, the computational complexity is reduced to $O(n^2 \cdot n_\mathrm{q})$. Nevertheless, the computation of the inverse is probably the most expensive one. This operation is limited by $O(n^3)$. Equation (8) serves as the reference for our proposed efficient computation of the diagonal of the redundancy matrix.

Premultypling Equation (6) with the material matrix yields

$$\mathbf{s} = -\mathbf{C}\mathbf{R}\mathbf{e}_0. \tag{9}$$

The symmetric matrix $\mathbf{C}\mathbf{R}$ is the so-called self-stress matrix, and Equation (9) is the map of the generalized pre-strains to the generalized stress resultants.

For further algebraic and spectral properties of the redundancy matrices in a discrete framework as well as redundancy functions in a continuous framework, we refer to the articles von Scheven et al. (2021) and Gade et al. (2021).

## 3    Method

In this section, a new formulation for the computation of redundancy matrices in large truss and frame structures is presented. According to the previous section, the costly computation of the redundancy matrix $\mathbf{R}$, see Equation (7), or its diagonal entries, see Equation (8), is dominated by inverting the $n \times n$ elastic stiffness matrix $\mathbf{K}$. This step becomes more and more challenging with a growing number of degrees of freedom $n$. From a mathematical point of view, however, the redundancy matrix has a rank equal to the degree of statical indeterminacy:

$$\mathrm{rank}(\mathbf{R}) = n_s = n_\mathrm{q} - n. \tag{10}$$

This difference compared to the total number of load-carrying modes in the structure can be defined as the relative degree of statical indeterminacy $\alpha = n_s/n_\mathrm{q} = (n_\mathrm{q} - n)/n_\mathrm{q}$, which satisfies $0 < \alpha \leqslant 1$. An analogous relative degree is known in adjustment of geodetic networks (Krarup 2006, p. 301). For small and moderate values of $\alpha$, the conventional computation of the redundancy matrix (involving the inverse of an $n \times n$ matrix) may be inefficient. Our method builds upon this fact and is described in the following subsections. In the first subsection, a decomposition of the redundancy matrix is introduced. Then, the subsequent subsection makes use of this decomposition and presents our algorithms to efficiently compute either the entire redundancy matrix or the diagonal of the redundancy matrix.

### 3.1    Decomposition of the redundancy matrix

While Pellegrino (1993) used the SVD of the equilibrium matrix $\mathbf{A}^\top$ to relate it to physical properties, we propose to perform the SVD on the matrix product $\mathbf{C}^{\frac{1}{2}}\mathbf{A}$ to obtain new insights into the computation of redundancy matrices. The matrix $\mathbf{C}^{\frac{1}{2}}\mathbf{A}$ is also considered within a matrix force method by Robinson and Haggenmacher (1970). Let the SVD of $\mathbf{C}^{\frac{1}{2}}\mathbf{A} \in \mathbb{R}^{n_\mathrm{q} \times n}$ be given by

$$\mathbf{C}^{\frac{1}{2}}\mathbf{A} = \mathbf{U}\mathbf{\Sigma}\mathbf{V}^\top, \tag{11}$$

where $\mathbf{U} = [\mathbf{u}_1, \ldots, \mathbf{u}_n, \mathbf{u}_{n+1}, \ldots, \mathbf{u}_{n_\mathrm{q}}] \in \mathbb{R}^{n_\mathrm{q} \times n_\mathrm{q}}$ and $\mathbf{V} = [\mathbf{v}_1, \ldots, \mathbf{v}_n] \in \mathbb{R}^{n \times n}$ are orthogonal matrices, and

$$\mathbf{\Sigma} = \begin{bmatrix} \mathbf{S} \\ \mathbf{0} \end{bmatrix} \in \mathbb{R}^{n_\mathrm{q} \times n} \tag{12}$$

is a diagonal matrix with $\mathbf{S} = \mathrm{diag}(\sigma_1, \ldots, \sigma_n) \in \mathbb{R}^{n \times n}$ and $\sigma_1, \ldots, \sigma_n > 0$. Mathematically, the first $n$ column vectors $\mathbf{u}_1, \ldots, \mathbf{u}_n$ of the matrix $\mathbf{U}$ belong to the image of the matrix product $\mathbf{C}^{\frac{1}{2}}\mathbf{A}$.

In contrast, the last $n_q - n$ column vectors $u_{n+1}, \ldots, u_{n_q}$ belong to the left kernel of the matrix product $C^{\frac{1}{2}}A$ or, in other words, to the kernel of $(C^{\frac{1}{2}}A)^\top$, i.e., for $k = n+1, \ldots, n_q$ holds:

$$(u_k^\top C^{\frac{1}{2}} A)^\top = (C^{\frac{1}{2}} A)^\top u_k = 0. \tag{13}$$

Now, we can integrate the above decomposition of $C^{\frac{1}{2}}A$ into the redundancy matrix formula. To do this, our first step is to rearrange the redundancy matrix equation in Equation (7) as follows:

$$
\begin{aligned}
R &= I - AK^{-1}A^\top C \\
&= I - A(A^\top CA)^{-1}A^\top C \\
&= I - C^{-\frac{1}{2}}C^{\frac{1}{2}}A(A^\top C^{\frac{1}{2}}C^{\frac{1}{2}}A)^{-1}A^\top C^{\frac{1}{2}}C^{\frac{1}{2}} \\
&= I - C^{-\frac{1}{2}}U\Sigma V^\top (V\Sigma^\top U^\top U\Sigma V^\top)^{-1}V\Sigma^\top U^\top C^{\frac{1}{2}}.
\end{aligned}
\tag{14}
$$

Since the matrices $U$ and $V$ are orthogonal, i.e., $U^\top = U^{-1}$ and $V^\top = V^{-1}$, the inverse in the above equation can be calculated via the following formula:

$$(V\Sigma^\top U^\top U\Sigma V^\top)^{-1} = V(\Sigma^\top \Sigma)^{-1}V^\top. \tag{15}$$

With this, we can rearrange the redundancy matrix equation via Equation (14) and Equation (15) further:

$$
\begin{aligned}
R &= I - C^{-\frac{1}{2}}U\Sigma V^\top (V\Sigma^\top U^\top U\Sigma V^\top)^{-1}V\Sigma^\top U^\top C^{\frac{1}{2}} \\
&= I - C^{-\frac{1}{2}}U\Sigma V^\top V(\Sigma^\top \Sigma)^{-1}V^\top V\Sigma^\top U^\top C^{\frac{1}{2}} \\
&= I - C^{-\frac{1}{2}}U\Sigma(\Sigma^\top \Sigma)^{-1}\Sigma^\top U^\top C^{\frac{1}{2}} \\
&= I - C^{-\frac{1}{2}}U \begin{bmatrix} S \\ 0 \end{bmatrix} \left( \begin{bmatrix} S & 0 \end{bmatrix} \begin{bmatrix} S \\ 0 \end{bmatrix} \right)^{-1} \begin{bmatrix} S & 0 \end{bmatrix} U^\top C^{\frac{1}{2}} \\
&= I - C^{-\frac{1}{2}}U \begin{bmatrix} S \\ 0 \end{bmatrix} (S^2)^{-1} \begin{bmatrix} S & 0 \end{bmatrix} U^\top C^{\frac{1}{2}} \\
&= I - C^{-\frac{1}{2}}U \begin{bmatrix} I \\ 0 \end{bmatrix} \begin{bmatrix} I & 0 \end{bmatrix} U^\top C^{\frac{1}{2}} \\
&= I - C^{-\frac{1}{2}}[u_1, \ldots, u_n][u_1, \ldots, u_n]^\top C^{\frac{1}{2}}.
\end{aligned}
\tag{16}
$$

The above equation reveals the mathematical matrix structure of the redundancy matrix. In fact, the conventional computation of the redundancy matrix translates into a computation of an orthogonal basis for the image of the matrix product $C^{\frac{1}{2}}A$. Both approaches mainly reflect a computational complexity that is characterized by the number of degrees of freedom $n$. However, our goal is to obtain a formula whose computational complexity is characterized by the difference $n_q - n$. To achieve this, we use the relation

$$I - [u_1, \ldots, u_n][u_1, \ldots, u_n]^\top = [u_{n+1}, \ldots, u_{n_q}][u_{n+1}, \ldots, u_{n_q}]^\top, \tag{17}$$

which reflects the connection between orthogonal projections (onto the image of $C^{\frac{1}{2}}A$ and kernel of $(C^{\frac{1}{2}}A)^\top$). In this way, we obtain our final result of the redundancy matrix using Equation (16) and Equation (17):

$$
\begin{aligned}
R &= I - C^{-\frac{1}{2}}[u_1, \ldots, u_n][u_1, \ldots, u_n]^\top C^{\frac{1}{2}} \\
&= C^{-\frac{1}{2}}C^{\frac{1}{2}} - C^{-\frac{1}{2}}[u_1, \ldots, u_n][u_1, \ldots, u_n]^\top C^{\frac{1}{2}} \\
&= C^{-\frac{1}{2}}(I - [u_1, \ldots, u_n][u_1, \ldots, u_n]^\top)C^{\frac{1}{2}} \\
&= C^{-\frac{1}{2}}([u_{n+1}, \ldots, u_{n_q}][u_{n+1}, \ldots, u_{n_q}]^\top)C^{\frac{1}{2}} \\
&= C^{-1}(C^{\frac{1}{2}}[u_{n+1}, \ldots, u_{n_q}])(C^{\frac{1}{2}}[u_{n+1}, \ldots, u_{n_q}])^\top.
\end{aligned}
\tag{18}
$$

The above final formula shows that the redundancy matrix can be computed via the $n_q - n$ column vectors $[u_{n+1}, \ldots, u_{n_q}]$ and simple algebraic manipulations (e.g., multiplication with a diagonal

```
1   function EfficientRedundancyMatrix(A, C)
2       Compute C^(1/2)A ∈ ℝ^(n_q × n)
3       Compute [u_{n+1}, ..., u_{n_q}] ∈ ℝ^(n_q × (n_q − n))          ▷ orthogonal basis for the kernel of (C^(1/2)A)^⊤
4       Compute C^(1/2)[u_{n+1}, ..., u_{n_q}] ∈ ℝ^(n_q × (n_q − n))
5       Compute (C^(1/2)[u_{n+1}, ..., u_{n_q}])(C^(1/2)[u_{n+1}, ..., u_{n_q}])^⊤ ∈ ℝ^(n_q × n_q)
6       Compute C^(−1)(C^(1/2)[u_{n+1}, ..., u_{n_q}])(C^(1/2)[u_{n+1}, ..., u_{n_q}])^⊤ ∈ ℝ^(n_q × n_q)          ▷ Equation (18)
7   end function
```

**Algorithm 1**    Efficient computation of the entire redundancy matrix in large truss or frame structures.

matrix and outer products). In particular, this computation has a computation complexity that is characterized by the difference $n_q - n$. Thus, the smaller the difference $n_q - n$ is, the more efficient the formula gets. The only numerical issue is to compute the vectors $[u_{n+1}, \ldots, u_{n_q}]$ efficiently, i.e., in other words, to efficiently compute an orthogonal basis for the kernel of $(C^{\frac{1}{2}}A)^\top$. This aspect is also discussed in the next subsection.

The column vectors $u_1, \ldots, u_n, u_{n+1}, \ldots, u_{n_q}$ characterize not only the kernel and image of $(C^{\frac{1}{2}}A)^\top$ but also the spectral components of the redundancy matrix. In fact, the columns of $C^{-\frac{1}{2}}[u_1, \ldots, u_n]$ are eigenvectors of $R$ to the eigenvalue 0 and the columns of $C^{-\frac{1}{2}}[u_{n+1}, \ldots, u_{n_q}]$ are the eigenvectors to the eigenvalue 1. To prove it, we use Equation (18) and consider the product of the redundancy matrix with an arbitrary column vector $C^{-\frac{1}{2}}u_l$, i.e.,

$$R(C^{-\frac{1}{2}}u_l) = C^{-\frac{1}{2}}[u_{n+1}, \ldots, u_{n_q}][u_{n+1}, \ldots, u_{n_q}]^\top C^{\frac{1}{2}}(C^{-\frac{1}{2}}u_l) \qquad (19)$$
$$= C^{-\frac{1}{2}}[u_{n+1}, \ldots, u_{n_q}][u_{n+1}, \ldots, u_{n_q}]^\top u_l$$
$$= \begin{cases} C^{-\frac{1}{2}}u_l & n+1 \le l \le n_q, \\ 0 & l \le n. \end{cases}$$

Another way of interpreting Equation (18) is to relate the formula to the self-stress matrix $CR$ (see Equation (9)). In fact, Equation (18) can be rearranged into

$$CR = (C^{\frac{1}{2}}[u_{n+1}, \ldots, u_{n_q}])(C^{\frac{1}{2}}[u_{n+1}, \ldots, u_{n_q}])^\top. \qquad (20)$$

This formula shows that the matrix $C^{\frac{1}{2}}[u_{n+1}, \ldots, u_{n_q}]$ decomposes the self-stress matrix $CR$. This decomposition is an intermediate result in Algorithm 1 (see Line 5). Therefore, our formula also covers the efficient computation of the self-stress matrix.

Analogously to the redundancy matrix, there is a relationship of the column vectors $u_1, \ldots, u_n, u_{n+1}, \ldots, u_{n_q}$ to the spectral components of the self-stress matrix. While the columns of $C^{-\frac{1}{2}}[u_1, \ldots, u_n]$ are eigenvectors of $CR$ to the eigenvalue 0, the columns of $C^{-\frac{1}{2}}[u_{n+1}, \ldots, u_{n_q}]$ solve the generalized eigenvalue problem of $CR$ relative to $C$ to the eigenvalue 1. To realize this, Equation (19) is multiplied with $C$ (from the left), i.e.,

$$CR(C^{-\frac{1}{2}}u_l) = \begin{cases} C(C^{-\frac{1}{2}}u_l) & n+1 \le l \le n_q, \\ 0 & l \le n, \end{cases} \qquad (21)$$

which proves the eigenvalue statement mentioned above.

### 3.2    Efficient computation of the redundancy matrix

In the previous subsection, the formula in Equation (18) for the redundancy matrix is presented that favours an efficient computation if the difference $n_q - n$ is small. Based on this formula, we derive two different algorithmic approaches to compute either the entire redundancy matrix or only the diagonal of the redundancy matrix. Both scenarios use the column vectors $[u_{n+1}, \ldots, u_{n_q}]$, which is an orthogonal basis for the kernel of $(C^{\frac{1}{2}}A)^\top$. This is the first step in both algorithms: Algorithms 1 and 2.

Numerically, the computation of an orthogonal basis for the kernel of $(C^{\frac{1}{2}}A)^\top$ can be performed via QR decomposition or another optimized function. Since our implementation was done with MATLAB, we propose the use of the built-in QR decomposition function via the command $[\mathbf{Q}, \sim, \sim] = \mathrm{qr}(\cdot)$ (with this notation, the QR decomposition makes use of permutation matrices to reduce fill-in). Even though the resulting orthogonal matrix $\mathbf{Q} = [\mathbf{u}_1, \ldots, \mathbf{u}_n, \mathbf{u}_{n+1}, \ldots, \mathbf{u}_{n_q}]$ also consists of the image of $(\mathbf{C}^{\frac{1}{2}} \mathbf{A})$ and the kernel of $(\mathbf{C}^{\frac{1}{2}} \mathbf{A})^\top$, to our knowledge, it is the most efficient method to extract the desired column vectors $[\mathbf{u}_{n+1}, \ldots, \mathbf{u}_{n_q}]$.

In the following, we present the further steps for each of the two scenarios. We start with the efficient computation of the entire redundancy matrix.

**Efficient computation of the entire redundancy matrix**    Algorithm 1 presents the efficient computation of the entire redundancy matrix. As explained previously, the computation of the column vectors $[\mathbf{u}_{n+1}, \ldots, \mathbf{u}_{n_q}]$ is the first step of the algorithm and is represented in Lines 2–3. The next three lines implement the main formula that is described in Equation (18): while Line 4 and 6 simply perform matrix multiplications with diagonal matrices, Line 5 is a sum of outer product.

The computational complexity of the algorithm is mainly characterized by the computation of an orthogonal basis for the kernel of $(\mathbf{C}^{\frac{1}{2}} \mathbf{A})^\top$ (as the other operations are very simple). Depending on the numerical method, the extraction of the $n_q - n$ vectors can be very efficient. This leads to a fast computation of the entire redundancy matrix for small and moderate values $\alpha$.

**Efficient computation of the diagonal of the redundancy matrix**    Algorithm 2 presents the efficient computation of the diagonal entries of the redundancy matrix. As explained in the beginning of this subsection, the computation of the column vectors $[\mathbf{u}_{n+1}, \ldots, \mathbf{u}_{n_q}]$ is the first step of the algorithm and represented in Lines 2–3. Since our goal is to only extract the diagonal entries of the redundancy matrix, we can rearrange Equation (18) as

$$
\begin{aligned}
\mathrm{diag}(\mathbf{R}) &= \mathrm{diag}(\mathbf{C}^{-1}(\mathbf{C}^{\frac{1}{2}}[\mathbf{u}_{n+1}, \ldots, \mathbf{u}_{n_q}])(\mathbf{C}^{\frac{1}{2}}[\mathbf{u}_{n+1}, \ldots, \mathbf{u}_{n_q}])^\top) \\
&= \mathrm{diag}(\mathbf{C}^{-\frac{1}{2}}[\mathbf{u}_{n+1}, \ldots, \mathbf{u}_{n_q}][\mathbf{u}_{n+1}, \ldots, \mathbf{u}_{n_q}]^\top \mathbf{C}^{\frac{1}{2}}) \\
&= \sum_{l=1}^{n_q} (\mathbf{e}_l^\top \mathbf{C}^{-\frac{1}{2}}[\mathbf{u}_{n+1}, \ldots, \mathbf{u}_{n_q}][\mathbf{u}_{n+1}, \ldots, \mathbf{u}_{n_q}]^\top \mathbf{C}^{\frac{1}{2}} \mathbf{e}_l) \cdot \mathbf{e}_l \\
&= \sum_{l=1}^{n_q} (\mathbf{e}_l^\top [\mathbf{u}_{n+1}, \ldots, \mathbf{u}_{n_q}][\mathbf{u}_{n+1}, \ldots, \mathbf{u}_{n_q}]^\top \mathbf{e}_l) \cdot \mathbf{e}_l \\
&= \begin{pmatrix} \sum_{k=n+1}^{n_q} \mathbf{u}_{k,1}^2 \\ \sum_{k=n+1}^{n_q} \mathbf{u}_{k,2}^2 \\ \vdots \\ \sum_{k=n+1}^{n_q} \mathbf{u}_{k,n_q}^2 \end{pmatrix},
\end{aligned}
\tag{22}
$$

where $\mathbf{e}_l$ is the $l$-th unit vector and $\mathbf{u}_{k,l}$ is the $l$-th entry of the $k$-th vector. The resulting formula is very simple and can be expressed in a more compact way, that is, the sum of each row of the matrix $[\mathbf{u}_{n+1}, \ldots, \mathbf{u}_{n_q}] \circ [\mathbf{u}_{n+1}, \ldots, \mathbf{u}_{n_q}]$, where $\circ$ is the Hadamard product (element-wise product). This procedure is the second step in Algorithm 2 and is described in Lines 4–5.

The computational complexity of the algorithm is dominated by the computation of an orthogonal basis for the kernel of $(\mathbf{C}^{\frac{1}{2}} \mathbf{A})^\top$ (as the other operations are almost negligible). Depending on the numerical method, the extraction of the $n_q - n$ vectors can be very efficient. This leads to an extremely fast computation of the diagonal entries of the redundancy matrix for small and moderate values of $\alpha$.

## 4   Examples

This section demonstrates the application of the proposed efficient computation formulations for both computations of the distribution of the degree of statical indeterminacy and computation of the full redundancy matrix. With the first example, using a generic truss structure, the dependency of the efficiency of the proposed formulations on the relative degree of statical indeterminacy $\alpha$ will be demonstrated. The last two examples will apply the efficient computation

```
1  function EFFICIENTREDUNDANCYMATRIXDIAGONAL(A, C)
2      Compute C^(1/2) A ∈ ℝ^(n_q × n)
3      Compute [u_{n+1}, ..., u_{n_q}] ∈ ℝ^(n_q × (n_q − n))        ▷ orthogonal basis for the kernel of (C^(1/2) A)^⊤
4      Compute Hadamard product [u_{n+1}, ..., u_{n_q}] ∘ [u_{n+1}, ..., u_{n_q}] ∈ ℝ^(n_q × (n_q − n))
5      Compute sum of each row of [u_{n+1}, ..., u_{n_q}] ∘ [u_{n+1}, ..., u_{n_q}]        ▷ Equation (22)
6  end function
```

**Algorithm 2**  Efficient computation of the distribution of the degree of statical indeterminacy (diagonal of the redundancy matrix) in large truss or frame structures.

to a more realistic truss structure and a three-dimensional frame system. All three examples are scalable to demonstrate the efficiency of the proposed algorithms for different problem sizes. For all examples, even a random distribution of the stiffnesses of the bars does not affect the increase in efficiency. All computations were done with MATLAB R2022a on a machine with a 2.80 GHz Intel Core i7-1165G7 processor and 32 GB RAM. An implementation of the algorithms and all examples is publicly available on DaRUS (Tkachuk et al. 2023).

## 4.1   Truss cylinder

The first example uses a cylindrical truss structure as it has been used for gas holders. The cylinder has a radius of 1 m and a height of 10 m. To examine structures with different numbers of elements and degrees of freedom, the number of segments in the axial and circumferential direction is defined by $n$ and varied between 5 and 100. Furthermore, the degree of statical indeterminacy is varied by using different numbers of diagonal bracings. The relative degree of statical indeterminacy $\alpha$ will thus vary between 0.1 and 0.4. Figure 1 shows three structures with $n = 8$ and $\alpha = 0.1, 0.25, 0.4$.

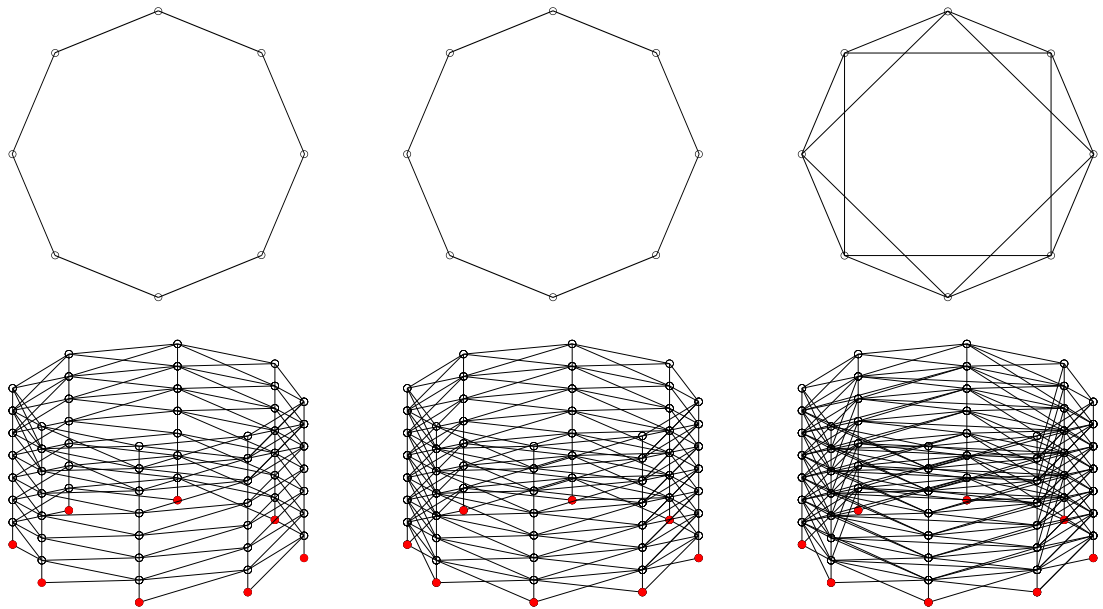

**Figure 1**  Truss cylinder: Systems with $n = 8$ and $\alpha = 0.1, 0.25, 0.4$. Top view and perspective view. For a better visualization, the vertical axis is scaled differently than the axes in the horizontal plane.

Young's modulus $E$ and cross-sectional area $A$ are constant for all elements. The nodes in the lower level (marked in red) are fixed in all three directions.

Three series of comparisons with different values of $\alpha$ were examined for both the computation of the full redundancy matrix and of the distribution of the degree of statical indeterminacy. For each series, $n$ was varied from 5 to 100 by steps of 5. But due to memory limitations on the reference machine, the series for the computation of the full redundancy matrix could not be completed up to $n = 100$.

In Figure 2, the results are shown for the computation of the full redundancy matrix. The left diagram shows the absolute computation time in seconds for the standard algorithms and the proposed efficient computation depending on the number of segments $n$ for the truss cylinder with $\alpha = 0.1$. It can be observed that the absolute gain in computation time grows with $n$. The relative speedup for all three values of $\alpha$ is shown in the right diagram. For $\alpha = 0.1$ (purple line), the speedup is in the range between 4 and 6 for most problem sizes.

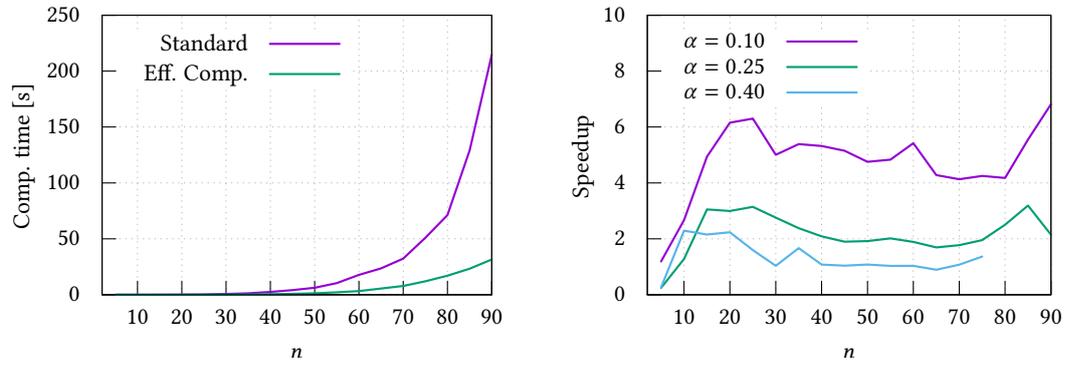

**Figure 2**  Truss cylinder: Computation times for $\alpha = 0.1$ and speedup for $\alpha = 0.1, 0.25, 0.4$ for computation of full redundancy matrix.

For the configurations with a higher relative degree of statical indeterminacy $\alpha$, only the speedup is shown in the right diagram. It can be observed that the speedup highly depends on the value of $\alpha$. For this truss system and $\alpha = 0.4$, the speedup goes down to values between 1 and 2. This means, the proposed efficient algorithm is still faster for all values of $n$, but the benefit is not as large as for systems with a lower relative degree of statical indeterminacy $\alpha$.

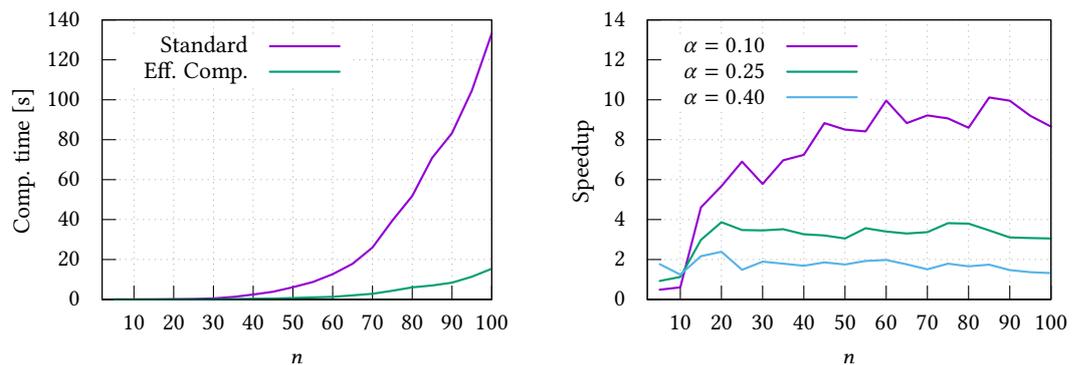

**Figure 3**  Truss cylinder: Computation times for $\alpha = 0.1$ and speedup for $\alpha = 0.1, 0.25, 0.4$ for computation of distribution of the degree of statical indeterminacy.

As a second scenario, we consider now the computation of the distribution of the degree of statical indeterminacy, i.e., only of the diagonal entries of the redundancy matrix. The proposed Algorithm 2 is compared against the reference method described by Equation (8). In Figure 3 the absolute computation times are visualized for the cylinder with $\alpha = 0.1$ (left) and the speedup for all three values of $\alpha$ (right). In general, we can observe a similar behavior as for the computation of the full redundancy matrix. The absolute gain in computation time grows with $n$ and the speedup decreases with the relative degree of statical indeterminacy $\alpha$. However, the speedup is generally at a higher level and even reaches values up to 10 for $\alpha = 0.1$.

## 4.2  Mero roof

The second example examines the calculation of the distribution of the degree of statical indeterminacy and the full redundancy matrix for a three-dimensional curved MERO space truss, as it is used in many buildings all over the world, e.g., in the Stockholm Globe Arena or the new Leipzig Trade Fair. The structure as shown in Figure 4 in two perspectives consists of two offset layers of square cells connected with diagonal bars. Each cell is $1\,\mathrm{m} \times 1\,\mathrm{m}$ in the $x - y$ plane. In this example, a quadratic function is used for the curvature in $x$ and $y$ direction. Young's modulus $E$ and cross-sectional area $A$ are constant for all elements. The nodes at the corners of the bottom layer (marked in red) are fixed in all three directions.

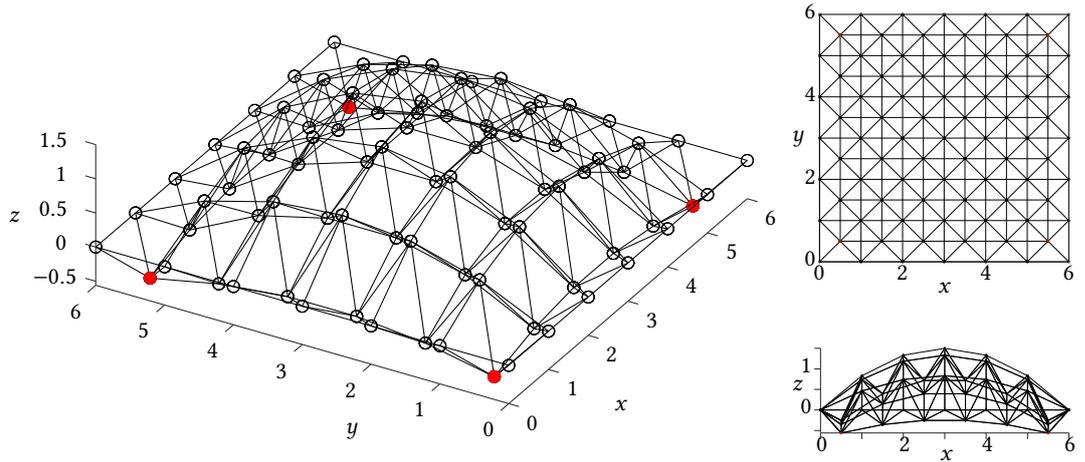

**Figure 4** Mero roof: System for $n = 6$.

For this structure the total degree of statical indeterminacy is equal to 45. The spatial distribution of the degree of statical indeterminacy in the structure is shown in Figure 5. Elements with a small contribution to the degree of statical indeterminacy are drawn with thin lines in light blue, while elements with a larger contribution are shown as thicker lines in darker blue. It can be observed that, e.g. the three bars connected to the top corners have a very small contribution to the degree of statical indeterminacy. In fact, they are statically determinate and therefore have a value of zero. In contrast, the bars in the lower layer generally have a higher contribution to the degree of statical indeterminacy as these bars undergo more constraint due to the four supports.

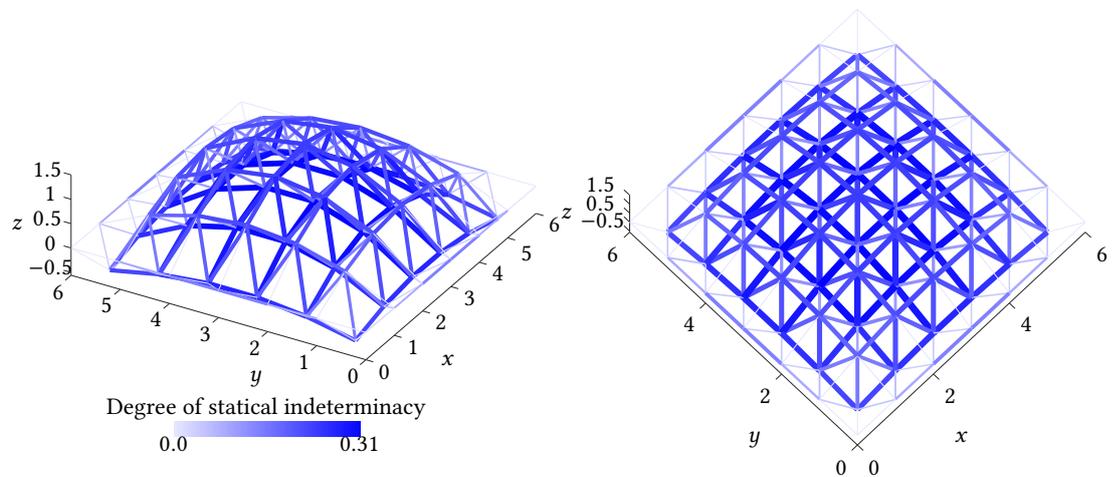

**Figure 5** Mero roof: Distribution of the degree of statical indeterminacy for $n = 6$.

To examine different problem sizes, the number of cells $n$ in $x$ and $y$ direction is varied between 5 and 85. Since the topology of the structure is given, the relative degree of statical indeterminacy cannot be varied. For larger values of $n$, $\alpha$ tends toward the value 0.24.

In Figure 6, the computation time and speedup are shown for the full redundancy matrix (left) and the distribution of the degree of statical indeterminacy (right). Due to memory limitations, the full redundancy matrix could only be calculated up to $n = 60$. For both cases, it can be observed that the computation time for the proposed efficient computation is much smaller compared to the standard computation. The speedup for the computation of the full redundancy matrix is around 3 and for the distribution of the degree of statical indeterminacy even up to 5 or 6.

This example shows that the proposed efficient computation allows a significant acceleration of the computation even for real spatial trusses with more than 50,000 elements.

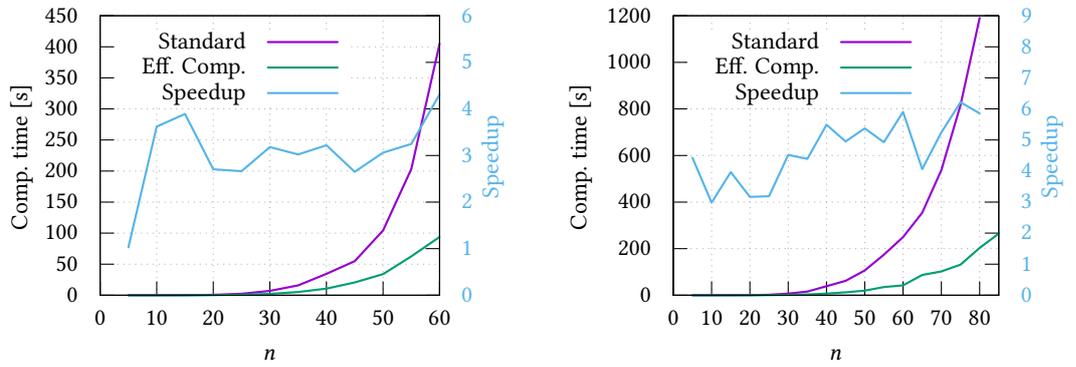

**Figure 6**   Mero roof: Computation times and speedup for computation of full redundancy matrix (left) and distribution of the degree of statical indeterminacy (right).

### 4.3   3D frame system

In the last example, efficient computation is applied to a three-dimensional frame structure. The structure shown in Figure 7 is a hyperbolic paraboloidal grid shell with square cells. Two neighboring edges (marked in red) are clamped to support the structure. All beams are rigidly connected and have the same Young's modulus $E$ and cross-section.

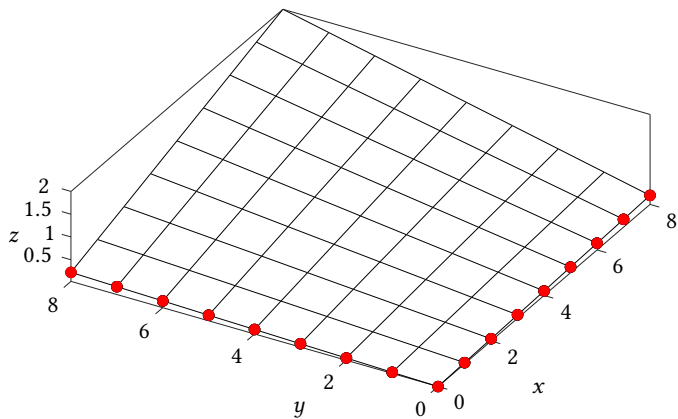

**Figure 7**   3D frame system: System for $n = 8$.

For spatial beam elements, the number of load-carrying modes is $n_m = 6$. Therefore, also the degree of statical indeterminacy for every element can be up to 6. For this hyperbolic paraboloidal grid shell, the relative degree of statical indeterminacy tends toward $\alpha = 0.5$ for large values of $n$.

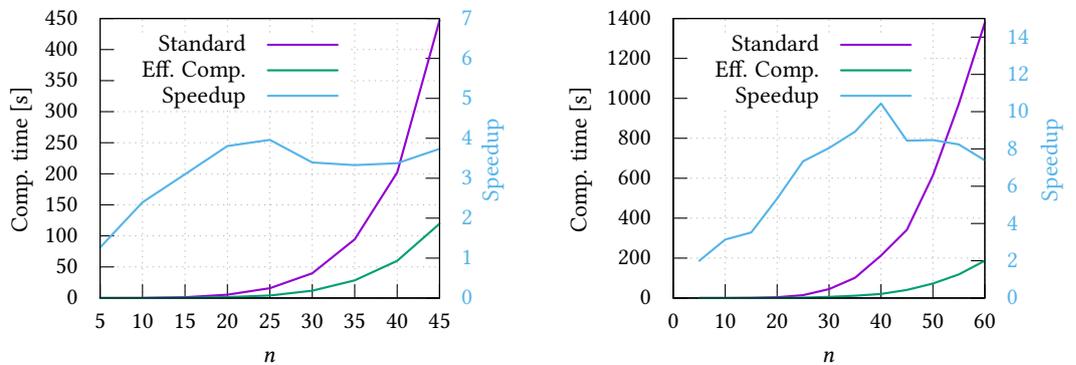

**Figure 8**   3D frame system: Computation times and speedup for computation of full redundancy matrix (left) and distribution of the degree of statical indeterminacy (right).

In Figure 8, the computation time and the speedup are shown for the full redundancy matrix (left) and the computation of the distribution of the degree of statical indeterminacy (right). Also, for this more complex mechanical model and a higher value of the relative degree of statical indeterminacy, a considerable speedup can be observed. For larger values of $n$, the speedup obtained by the proposed efficient computation for the distribution of the degree of statical indeterminacy is around 8 and for the efficient computation of the full redundancy matrix still between 3 and 4.

## 5 Conclusion

The paper addresses the problem of efficient computation of redundancy matrices for large truss and frame structures. Our proposed new formula is based on a SVD decomposition of the matrix product between the square root of the material matrix and the compatibility matrix. The few last columns of the decomposition construct the eigenspace of the redundancy matrix and provide a compact expression for the redundancy matrix. Our experiments have shown that for moderate values of a relative degree of statical indeterminacy, i.e., around $\alpha = 0.4$, a speedup is observed for the full redundancy matrix computation. For lower values of $\alpha$, our proposed method allows for much higher speedups (up to 5 times faster). This speedup is stable over the number of members in a structure, as the three considered examples with scalable structures illustrate. Even a more significant speedup (up to 10 times faster) is observed for the computation of the distribution of the degree of statical indeterminacy, which is given by diagonal entries of the redundancy matrix.

The current study is limited to computations on CPU and straight structural members. In the future, hardware acceleration for the new computation algorithms on GPU may be considered. Curved beams are usually discretized with elements with more nodes and larger bandwidths. Thus, the behavior of the proposed algorithm may differ for frames with curved members. Finally, a new update algorithm for the full redundancy matrix based on our proposed decomposition can be developed and compared with the existing update algorithm from Krake et al. (2022). The implementation of the update should be straightforward, e.g., it profits from existing algorithms for dense and sparse matrices (Reichel and Gragg 1990; Andrew and Dingle 2014).

## A  Factorization of the stiffness matrix for a 3D Euler-Bernoulli beam element

Herein, the expression from (Przemieniecki 1968) for the elastic stiffness matrix of a 2-node straight Euler-Bernoulli beam in 3D with a constant cross-section along the length is used. The local coordinate system $\{\hat{e}_i\}_{i=1}^{3}$ is aligned with the axial direction for $i = 1$ and the principle bending axes $i = 2, 3$ with the second moments of inertia $I_{yy}$ and $I_{zz}$, respectively. A factorization of the elastic stiffness matrix $\mathbf{K}_e$ with a diagonal material matrix reads

$$\mathbf{C}_e = \frac{1}{L} \operatorname{diag} \left\{ \begin{bmatrix} EA & GJ & 3EI_{zz} & EI_{zz} & 3EI_{yy} & EI_{yy} \end{bmatrix} \right\}, \tag{A.1}$$

$$\mathbf{A}_e = \begin{bmatrix} -\hat{e}_1 & \mathbf{0} & 2\hat{e}_2/L & \mathbf{0} & -2\hat{e}_3/L & \mathbf{0} \\ \mathbf{0} & -\hat{e}_1 & \hat{e}_3 & -\hat{e}_3 & \hat{e}_2 & -\hat{e}_2 \\ \hat{e}_1 & \mathbf{0} & -2\hat{e}_2/L & \mathbf{0} & 2\hat{e}_3/L & \mathbf{0} \\ \mathbf{0} & \hat{e}_1 & \hat{e}_3 & \hat{e}_3 & \hat{e}_2 & \hat{e}_2 \end{bmatrix}, \tag{A.2}$$

where $A$ is the cross-sectional area, $J$ is the torsional moment of inertia, $E$ and $G$ are Young's and shear moduli, respectively. $\mathbf{0}$ in the latter expression are zero column-vectors $\mathbf{0}_{3\times1}$. The load-carrying modes are axial tension/compression, torsion, and two bending/shear modes along each principle bending direction. As described in (von Scheven et al. 2021, pp. 45/46), the normalization is included in the matrix $\mathbf{C}_e$, so that $\mathbf{K}_e = \mathbf{A}_e^\top \mathbf{C}_e \mathbf{A}_e$. This ensures a redundancy matrix that is independent of the chosen unit for the length dimension.

**Authors' contributions**    AT: Conceptualization, Software, Writing, Supervision; TK: Conceptualization, Methodology, Software, Writing; JG: Conceptualization, Software, Writing; MvS: Conceptualization, Software, Writing, Visualization

**Supplementary Material**    Demo files demonstrating the speedup of the proposed algorithms available at the permalink 10.18419/darus-3347.

**Acknowledgements**    Financial support of Anton TKACHUK by the Department of Engineering and Physics at Karlstad University via the internal program VG 5 is gratefully acknowledged.

**Funding**    This work is partly funded by "Deutsche Forschungsgemeinschaft" (DFG, German Research Foundation) – Project-ID 251654672 – TRR 161; Project-ID 279064222 – SFB 1244; and under Germany's Excellence Strategy – EXC 2120/1 – 390831618.

**Ethics approval and consent to participate**    Not applicable.

**Consent for publication**    Not applicable.

**Competing interests**    The authors declare that they have no competing interests.

**Journal's Note**    JTCAM remains neutral with regard to the content of the publication and institutional affiliations.